\documentclass[reprint,twocolumn,superscriptaddress,showpacs,nofootinbib,notitlepage]{revtex4-2}

\usepackage{graphicx}
\usepackage{latexsym,amsmath,amssymb,lmodern,float,url}
\usepackage{natbib}
\usepackage{xcolor}
\usepackage{microtype}
\usepackage{comment}

\usepackage[colorlinks=true,backref=false, linktocpage=true,
citecolor=blue,urlcolor=blue,linkcolor=blue,pdfpagemode=UseOutlines]{hyperref}

\def\Tr{\operatorname{{Tr}}}

\newcommand{\eq}[1]{Eq.~(\ref{#1})}

\begin{document}
\title{Semidefinite Programs at Finite Fermion Density}

\author{Scott Lawrence}
\email{scott.lawrence-1@colorado.edu}
\affiliation{Department of Physics, University of Colorado, Boulder, CO 80309, USA}
\date{\today}

%%%%%%%%%%%%%%%%%%%%%%%%%%%%%%%%%%%%%%%%%%%%%%%%%%%%%%%
\begin{abstract}
Semidefinite programs can be constructed to provide a non-perturbative view of the zero-temperature behavior of quantum systems. Past work has found that small semidefinite programs (relative to the dimension of Hilbert space) yield quantitatively accurate estimates for the ground-state energy and expectation values. This paper examines the properties of these semidefinite programs when applied to lattice-regulated field theories exhibiting fermion sign problems. Specifically on the finite-density Thirring model, small semidefinite programs still yield quantitatively accurate results, and there is no indication that these methods encounter exponential costs (analogous to the fermion sign problem) at finite chemical potential.
\end{abstract}

\maketitle

%%%%%%%%%%%%%%%%%%%%%%%%%%%%%%%%%%%%%%%%%%%%%%%%%%%%%%%
\section{Introduction}\label{sec:introduction}

Monte Carlo methods, in the framework of lattice field theory, are often the only technique available for nonperturbative computations of expectation values in quantum field theories. These methods generally proceed by transforming a quantum expectation value to an equivalent classical expectation value in one higher space-time dimension. Unfortunately, for a wide variety of systems (notably including relativistic fermions at finite density and the Hubbard model away from half-filling), the resulting classical expression possesses a Boltzmann factor which is generally complex, and therefore cannot be viewed as a probability distribution for Monte Carlo sampling. Under standard complexity-theoretic assumptions, a theorem of Troyer and Wiese states that there is no fully general solution to these \emph{fermion sign problems}~\cite{troyer2005computational}.

Despite the nonexistence of a general solution to fermion sign problems, individual systems may be analyzed with a variety of approaches: 
complex Langevin~\cite{Aarts:2008rr}, the density of states method~\cite{Langfeld:2016mct}, canonical methods~\cite{Alexandru:2005ix,deForcrand:2006ec}, reweighting methods~\cite{Fodor:2001au}, series expansions in the chemical potential~\cite{Allton:2002zi}, fermion bags~\cite{Chandrasekharan:2013rpa}, analytic continuation from imaginary chemical potentials~\cite{deForcrand:2006pv}, and contour deformation methods~\cite{Alexandru:2020wrj}. These methods are frequently successful for low-dimensional (typically $1+1$-dimensional) systems, but none have yet been able to provide reliable information about e.g.~cold, dense quark matter.

A separate family of approaches to the zero-temperature behavior of quantum systems has recently been developed: the \emph{quantum-mechanical bootstrap}~\cite{Han:2020bkb,Berenstein:2021dyf,Berenstein:2021loy,Nancarrow:2022wdr}. Here we use the observation that, for any quantum mechanical operator $\mathcal O$, the expectation value $\Tr \rho \mathcal O^\dagger \mathcal O$ must be non-negative in any state $\rho$. As a direct consequence, if we select a basis of operators $\{\mathcal O_i\}$ and a particular state $\rho$, we may construct a matrix guaranteed to be positive-semidefinite\footnote{Here and throughout, the notation $M \succeq 0$ is used to denote that a matrix is positive-semidefinite.}:
\begin{equation}\label{psd}
M_{ij} \equiv
\langle \mathcal O_i^\dagger \mathcal O_j
\rangle
\succeq 0
\text.
\end{equation}
That is, for any vector $v$, we have $v^T M v \ge 0$. When the chosen basis of operators is complete---when any operator in the Hilbert space can be obtained as a linear combination of the elements of the basis---any set of expectation values consistent with \eq{psd} and the system's commutation relations is obtainable by some quantum state. In other words, every possible inequality describing the quantum system is captured by \eq{psd}.

This bound by itself is not usually sufficient to provide interesting information about the ground state: after all, high-temperature states (and indeed far-from-equilibrium states) necessarily obey this bound as well. To do better we must include information about the Hamiltonian of interest.
For a chosen Hamiltonian $H$, the ground state $|\Omega\rangle$ is by definition the state that minimizes $\langle \Omega|H|\Omega\rangle \equiv E_0$. Because the above bound captures every possible inequality, the relation $\Tr \rho H \ge E_0$ can be derived from that bound.

The computational strategy of the quantum-mechanical bootstrap is to use \eq{psd} to derive lower bounds on the expectation value $\langle H\rangle$, which must be obeyed by all states. These are bounds on the ground-state energy, and as more operators are used in the construction of $M$, the bounds converge to the true ground state energy. Concretely, we can cast the task of finding the ground state energy as an optimization problem over the space of expectation values, as follows. For a given basis of operators $\mathcal O_i$, we consider all expectation values of the form $\langle \mathcal O_i^\dagger \mathcal O_j\rangle$. This defines the positive semidefinite matrix $M_{ij}$ of \eq{psd}. The commutation relations furnish a set of linear constraints on the matrix elements of $M$---that is, a sequence of matrices $C^{(k)}$ such that we require
\begin{equation}
0 = \Tr C^{(k)} M
\text.
\end{equation}
We also impose that the expectation value of the identity operator is $1$. This is equivalent to enforcing the correct normalization of the quantum states being looked at. Note that all of these constraints are linear in the expectation values, which is the space in which our optimization problem takes place.

Finally, as long as the Hamiltonian is obtainable by a linear combination of our chosen basis, there is a matrix $G$ such that $\Tr G M$ represents the expectation value of the Hamiltonian. We seek to minimize this linear combination of elements of $M$ subject to the above constraints. Because this optimization problem is convex, the minimum obtained is guaranteed to be the global minimum, and therefore a lower bound on $\langle H \rangle$ that is valid for all states.

This type of optimization problem is termed a semi-definite program. This method provides a rigorous lower bound on the ground state energy. As a result, this method is essentially dual to variational methods, which provide similarly rigorous upper bounds on the ground state energy. The combination is particularly powerful when applied to systems for which the ground state energy is a quantity of central physical concern---for example in the study of quantum-mechanical bound states.

Separate from the explicit bound on the ground-state energy, we also obtain estimates of all expectation values used in the construction of the SDP. In situations where the ground-state energy is not of direct physical interest (most relevantly here, lattice-regularized field theories), these estimates are central to the utility of SDP-based methods.

When the basis of operators is complete, performing this minimization requires manipulating objects of size exponential in the volume of the physical system being studied. The same is true of attempting to exactly diagonalize the Hamiltonian, and so SDP-based methods appear to have no advantage. However, in practice it has been found that operator bases that are far from complete nevertheless result in remarkably tight bounds on simple quantum systems. This was reported early on for few-matrix mechanics~\cite{Han:2020bkb} and various one-body systems~\cite{Berenstein:2021dyf}. The power of the quantum-mechanical bootstrap derives from this observation: it is often possible to construct a polynomial-sized truncated basis for which the bootstrapped bound and many inferred expectation values are close to their true values.

Semi-definite programs have played an important role in quantum field theory before the advent of the quantum-mechanical bootstrap being used in this work. Most prominently, the conformal bootstrap provides the most precise constraints available so far on the critical exponents of many conformal field theories~\cite{Kos:2016ysd,Simmons-Duffin:2015qma,Rattazzi:2008pe} (see~\cite{Poland:2018epd} for a review of these methods). Bootstrap methods applied to the S-matrix, historically important~\cite{chew1961s}, have also been recently revived~\cite{He:2018uxa,Caron-Huot:2020cmc}.

This work applies SDP-based methods to quantum field theories by first regularizing the field theories on a spatial lattice (in the Hamiltonian formalism). The resulting system is a (many-body) quantum-mechanical system to which the methods of ~\cite{Han:2020bkb,Berenstein:2021dyf,Berenstein:2021loy} are directly applicable. A similar approach was first used in~\cite{barthel2012solving} to treat the Hubbard model, and demonstrated directly on field theories in~\cite{Lawrence:2021msm}. The hierarchy of positive-definiteness constraints so central to all quantum mechanical bootstrap methods is known as the Navascues-Pironio-Acin hierarchy~\cite{navascues2007bounding,navascues2008convergent}.

The remainder of the paper is organized as follows. Section~\ref{sec:model} introduces the Thirring model, a frequent test bed of methods targetting the fermion sign problem. The algorithms used to construct semi-definite programs, and extract from them expectation values, are detailed in Section~\ref{sec:algorithm}. Because the SDPs related to a space of expectation values, they can be constructed either at finite volumes or directly at infinite volume. The convergence behavior at finite volume, where direct diagonalization of the Hamiltonian is possible for comparison purposes, is examined in Section~\ref{sec:convergence}. Results at infinite volume are presented in Section~\ref{sec:infinite}. Finally, Section~\ref{sec:discussion} discusses various aspects of the performance of this method, lays out open questions, and suggests avenues for future work.

All code used for this paper is available online~\cite{code}. SDPs are solved using the MOSEK toolkit~\cite{mosek}.

%%%%%%%%%%%%%%%%%%%%%%%%%%%%%%%%%%%%%%%%%%%%%%%%%%%%%%%
\section{Thirring Model}\label{sec:model}
The Thirring model is one of many afflicted by the fermion sign problem. Variants of this model have long been used as testbeds for methods to tackle sign problems~\cite{Alexandru:2015xva,Alexandru:2015sua,Alexandru:2017czx,Alexandru:2018fqp,Aarts:2008rr,Fujii:2015vha,Pawlowski:2013pje}. This work is in keeping with that tradition.

A lattice Thirring model with staggered fermions is defined by the Hamiltonian
\begin{widetext}
\begin{equation}\label{hamiltonian}
H_{\mathrm{Thirring}} = \sum_x \left[(-1)^x m c^\dag(x) c(x) + \mu c^\dag(x) c(x)\right] + \sum_{\langle x y\rangle} \Big[(-1)^x \frac{c^\dag(x) c(y) + c^\dag(y) c(x)}{2} - g^2 c(x) c^\dag(x) c^\dag(y) c(y)\Big]\text.
\end{equation}
\end{widetext}
Here the second sum is taken over all pairs of adjacent sites on an $L$-site lattice with periodic boundary conditions.
The operators $c(x)$ are defined to obey anticommutation relations according to
\begin{equation}
\{c(x), c(y)\} = 0\text{, and }
\{c^\dagger(x),c(y)\} = \delta_{x,y}
\text.
\end{equation}
 The lattice parameters $m$, $\mu$, and $g^2$ are the mass, chemical potential, and coupling respectively. In the continuum limit, this model has emergent Lorentz invariance, and describes a single two-component fermionic field.

\begin{figure}
\centering\includegraphics[width=0.9\linewidth]{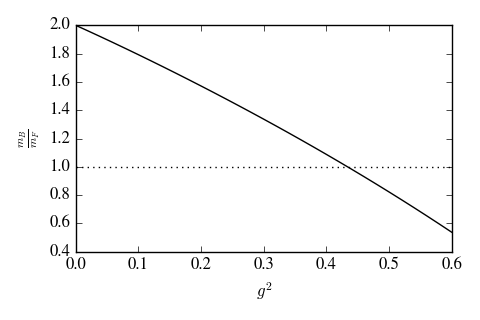}
\caption{Ratio of the fermion mass to the boson mass across couplings. The fermion mass is tuned to $m_F = 0.2$ in all cases, on a $10$-site lattice. The dotted gray line marks the point at which $m_B = m_F$, and heuristically, the crossing between the strong and weak coupling regimes.\label{fig:masses}}
\end{figure}

For positive $g^2$, this model has a repulsive interaction between fermions (attractive between a fermion and antifermion), producing a fermion-antifermion bound state analogous to mesons in quantum chromodynamics. The mass ratio $\frac {m_B}{m_F}$ of the boson to the meson may be taken as a measure of the strength of the coupling---at weak coupling $m_B \sim 2 m_F$, and studies of the strong-coupling sign problem typically aim for $m_B \sim m_F$~\cite{Alexandru:2016ejd}. This ratio $\frac{m_B}{m_F}$ is shown in Figure~\ref{fig:masses} across a range of couplings, with the fermion mass always tuned to be $m_F = 0.2$ in lattice units, and the calculation performed by exact diagonalization of the Hamiltonian on a $10$-site lattice. Calculations in this paper will typically be performed at $g^2=0.5$, well into the strong-coupling regime.

%%%%%%%%%%%%%%%%%%%%%%%%%%%%%%%%%%%%%%%%%%%%%%%%%%%%%%%
\section{Algorithm}\label{sec:algorithm}
The basic algorithm to study a quantum system via a semidefinite program proceeds in four steps. First, a basis of operators $\{\mathcal O_i\}$ is chosen. We will study the quantum system only by looking at expectation values of the form $\langle \mathcal O_i^\dagger \mathcal O_j\rangle$; all other properties are to be ignored.

The axioms of our quantum system imply two sets of facts about these expectation values. First, the positive-definiteness of the inner product on the Hilbert space implies that the matrix $M_{ij} \equiv \langle \mathcal O^\dagger_i \mathcal O_j\rangle$ is positive semi-definite. Second, the commutations relations imply certain linear relations between the expectation values. For example, for any sites $x,y$, we have $\langle c^\dagger(x) c(y) \rangle = \delta_{xy} - \langle c(y) c^\dagger(x) \rangle$. In the second step of the algorithm, these two sets of constraints are combined to construct a semi-definite program. The quantity to be minimized is the expectation value of the Hamiltonian, which can be expressed as a linear combination of matrix elements of $M$. The constraints are the linear relations between elements of $M$ derived from commutation relations, and the nonlinear requirement $M \succeq 0$.

Once the SDP has been constructed, the third step of the algorithm is to perform the constrained minimization. For this work, the minimization is done using the MOSEK toolkit~\cite{mosek}, which implements an interior-point method to perform the optimization~\cite{alizadeh1995interior}.

Finally, in the fourth step we wish to extract physical information from the solution of the SDP. For a simple quantum-mechanical system, the lower bound on the ground state is of intrinsic interest. In the context of (lattice-regularized) field theories, the ground state energy density at a single value of $\mu$ has no physically meaningful continuum limit. For zero-temperature, finite-chemical-potential models, the observable of greatest interest is typically the number density $n(\mu)$. Alternatively, the difference between ground-state energies\footnote{Note that in a relativistic theory, care must be taken to define $E(\mu)$ to exclude the contribution from the Fermi sea.} at two different values of $\mu$ provides the integral of $n(\mu)$ according to
\begin{equation}
E(\mu_i) - E(\mu_f) = \int_{\mu_i}^{\mu_f} n(\mu)
\text.
\end{equation}

When solving an SDP, typical algorithms (including the interior-point methods used by MOSEK) return not only the optimal value of the objective function (in this case corresponding to the ground-state energy), but also a value of the positive semi-definite matrix $M$ that obtains that optimum. Any observable encoded in the matrix $M$ can therefore be directly estimated once the optimization is complete. In the case of the Thirring model, because a term $\mu \hat N$ appears in the Hamiltonian, the fermion number will always be accessible from the expectation values present in $M$.

The fermion number could also be obtained, in principle, from solving two SDPs for slightly different values of the chemical potential $\mu$, and computing $\langle \hat N \rangle = -\frac{\partial}{\partial\mu}\langle \hat H - \mu \hat N\rangle$. In fact this yields the same result as simply reading off the value of $\langle \hat N\rangle$ from the matrix $M$ in the solution to the SDP. To see this, consider the positive semi-definite matrix $M$ obtained after solution. Generically, this matrix will have exactly one vanishing eigenvalue. The corresponding eigenvector $v$ defines some operator $\hat B \equiv v_i \mathcal O_i$ such that the expectation value $\langle \hat B^\dagger \hat B \rangle$ is estimated to be $0$. Using the linear part of the SDP (i.e.~the commutation relations), it may be shown that $\hat B^\dagger \hat B = \hat H - E$, where $E$ is the SDP-obtained estimate of the ground-state energy. Now let us consider what happens after perturbing the Hamiltonian by a term $\epsilon \hat N$. Again generically, $\hat B$ is also perturbed by some term of the same order $\hat B \rightarrow \hat B + \epsilon \hat C$. We now see that (expanding to first order in $\epsilon$)
\begin{equation}
\hat B^\dagger \hat B + \epsilon (\hat B^\dagger C + \hat C^\dagger B) = \hat H + \epsilon \hat N - E
\text,
\end{equation}
where again the equality $\hat B^\dagger \hat C + \hat C^\dagger \hat B = \hat N$ follows from the commutation relations. As a result, the change in the estimated ground state energy is equal to the expectation value of this operator, which is constrained by the commutation relations to be equal to the estimate $\langle \hat N \rangle$.

Typical computational methods (e.g.~lattice Monte Carlo methods) arrive at the infinite-volume limit by performing calculations at a sequence of finite volumes and then extrapolating to obtain the infinite-volume result. As discussed in~\cite{Lawrence:2021msm}, because SDP-based methods work directly with operators and make no reference to the Hilbert space itself, it is natural to perform calculations directly in the infinite-volume limit. This is accomplished with two modifications to the algorithm described above. First, instead of minimizing the Hamiltonian expectation value (which is no longer well-defined), we minimize the Hamiltonian density. Second, we impose translational invariance\footnote{Note that this embeds an additional assumption, namely that translational symmetry is not spontaneously broken in the finite-density ground state of the Thirring model.} by including in the formulation of the SDP additional linear constraints of the form $\langle \mathcal O \rangle = \langle T^\dagger \mathcal O T \rangle$ for any unitary translation operator $T$.

The core of the quantum-mechanical bootstrap is the choice of an incomplete basis of operators---in many ways this parallels the need for a choice of basis when explicitly diagonalizing a truncated Hamiltonian. The time required to solve an SDP scales roughly as the cube of the size of the basis, so it is critical to keep the basis small. At the same time, a poor choice of basis (including an insufficiently large basis) will not yield a tight bound on the ground-state energy. Finally note that, although the approximation to $E_0$ obtained via a hierarchy of SDPs is systematically improveable in the sense that, with a sufficiently large basis, the approximated $E_0$ can be brought within any desired tolerance of the exact result, in principle the cost may be exponential in the precision requested. An efficient algorithm depends on a well-chosen basis.

For this paper, we define a hierarchy of four bases. Each basis is both translationally invariant and Hermitian (meaning that for any operator $\mathcal O$ in the basis, $\mathcal O^\dagger$ is also present). The first, labelled H0, consists of $c(x)$, $c^\dagger(x)$, and $c^\dagger(x) c(x)$ for all lattice sites $x$. On an $L$-site lattice, counting the identity, this has $3L+1$ elements. Note that although the Hamiltonian is not in the span of this basis, it is in the span of operators obtained by taking a product $\mathcal O^\dagger_i \mathcal O_j$ of two elements of this basis, and so the SDP is able to provide a nontrivial constraint on the ground-state energy.

\begin{figure*}
\centering{
\includegraphics[width=0.45\linewidth]{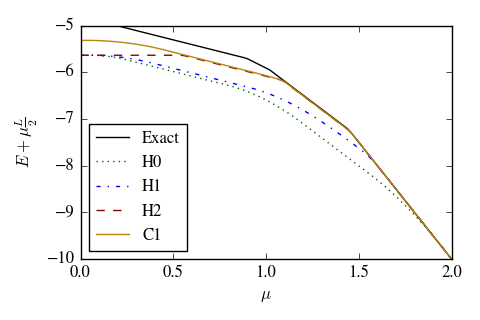}
\hfil
\includegraphics[width=0.45\linewidth]{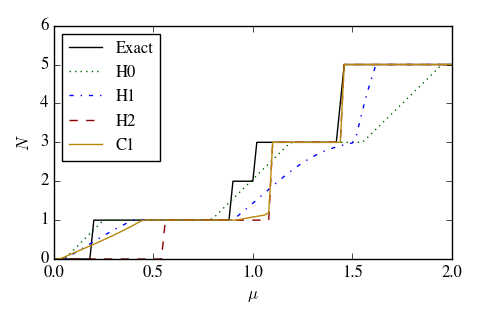}
}
\caption{SDP estimation of ground-state energy and fermion number for the Thirring model on $10$ sites, with $m=0.05$ and $g^2 = 0.5$, across the full range of chemical potentials. The left panel shows the estimate of the ground-state energy; the right compares the estimates of fermion number. The four bases used are described in the text.\label{fig:check}}
\end{figure*}

The next basis, H1, adds to H0 the $L$ operators of the form $c(x) c^\dagger(x+1)$, and their $L$ complex conjugates. This basis has a total of $5L+1$ elements.

To the elements of H1 we can add the $L$ operators found in the interaction term of the Hamiltonian: for each site $x$, $c^\dagger(x) c(x) c(x+1) c^\dagger(x+1)$. The basis obtained in this way is labelled H2.

These first three bases are of course inspired by terms found in the Hamiltonian of \eq{hamiltonian}. The final basis, termed C1, adds to H2 the $L$ operators of the form $c^\dagger(x) c(x+1) c^\dagger(x+2) c(x+2)$, and their complex conjugates. The size of this largest basis is $8L + 1$.

To check the correctness of these methods, Figure~\ref{fig:check} provides a comparison on a $10$-site lattice with ground state energy information obtained by exact diagonalization of the Hamiltonian. The left panel shows the ground-state energy bound proven via SDP. Although the bound itself is not meaningful, an estimate of the integral of number density between two chemical potentials may be obtained from the difference of energies at those values of $\mu$. (The extra term proportional to $\mu$ ensures that the Fermi sea is not counted in the estimated fermion number.) Alternatively, the right panel compares the estimated fermion number to the exact result. For all four bases, the energy bounds are seen to lie below (or on) the exact value, as they should.

Moreover, significant convergence towards the true value is seen as the basis is expanded. On this lattice, the largest basis (C1) has 81 elements. By comparison, a complete basis of operators on the $1024$ states would require $2^{20}$ operators.

The next section will consider the convergence properties of the SDPs obtained from these bases more carefully. Already from Figure~\ref{fig:check}, we can see that the basis H0 provides a qualitatively accurate equation of state everywhere away from the step at $\mu \sim 1.8$. The mechanism for this is visible in the left panel of that figure: for $\mu \in [0,1.5)$, the error in the energy is roughly constant.

%%%%%%%%%%%%%%%%%%%%%%%%%%%%%%%%%%%%%%%%%%%%%%%%%%%%%%%
\section{Convergence}\label{sec:convergence}
It has previously been demonstrated that semidefinite programs can yield efficient and accurate approximations for the ground-state properties of quantum lattices~\cite{barthel2012solving,Lawrence:2021msm}. Other methods, notably lattice Monte Carlo methods, suffer at finite chemical potentials and strong couplings due to the fermion sign problem. The purpose of this section is to investigate to what extent the same is true of the SDPs defined in the previous section.

\begin{figure*}
\centering{
\includegraphics[width=0.45\linewidth]{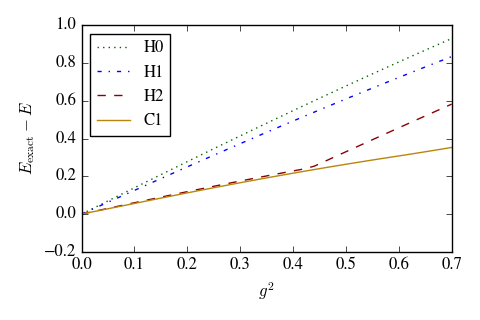}
\hfil
\includegraphics[width=0.45\linewidth]{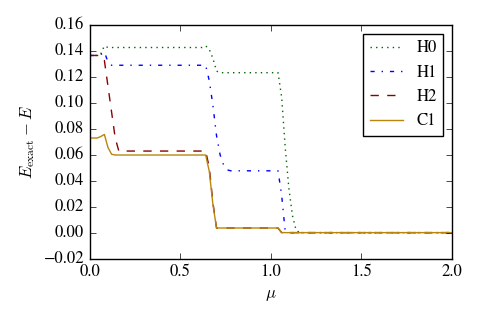}
}
\caption{Convergence of SDPs constructed from the four bases described in the text, across chemical potentials and couplings. The left panel shows the error in the ground state energy estimate as $g^2$ is changed, on $10$-site lattices, with the bare mass tunes to be $m_F=0.2$, and at $\mu=0.5$. All calculations in the right panel are performed at $g^2=0.5$ on a $10$-site lattice, with a bare mass of $m=0.05$.\label{fig:convergence}}
\end{figure*}

Figure~\ref{fig:convergence} shows the error in the ground-state energy estimate for all four bases described above. The most striking feature of this figure is that all SDPs perform best at \emph{large} chemical potentials---this is the opposite of the `traditional' behavior of Monte Carlo methods. This success is due to the fact that at large chemical potential, the true ground state of the system is saturated, with all fermionic modes being occupied. Any SDP will, at sufficiently large $\mu$, prefer this state to all others and therefore agree with the exact result.

As more operators are added to the basis, the SDP approximation improves. Again, the largest and easiest improvements are achieved at larger chemical potentials. In fact, of the bases considered, only the largest (C1) is substantially different at $\mu = 0$.

The news is not all good. Early attempts to resolve the fermion sign problem of lattice QCD and related models frequently struggled to replicate the so-called \emph{silver blaze phenomenon}~\cite{Cohen:2003kd}. (Contour deformations methods---such as those based on Lefschetz thimbles---were a notable exception, e.g.~\cite{Alexandru:2018ngw}.) In the infinite-volume limit, this term refers to the fact that the number density remains constant for some interval of chemical potentials $\mu \in [0,\mu_c)$ before suddenly beginning to increase. In QCD, this increase begins slightly below the proton mass---faulty algorithms for alleviating the sign problem would characteristically begin to increase far too early. We see that the SDPs investigated here mostly (H2 is a striking exception) have a similar flaw: the density begins to increase near the \emph{bare} fermion mass of $0.05$, rather than the correct physical mass $m_F \approx 0.20$.

Moreover, at a finite volume, all increases in the fermion number ought to be discontinuous jumps: the fermion number in a non-degenerate eigenstate of the Hamiltonian is always an integer. The semidefinite programs do not generally have this behavior, instead appearing to smooth out these jumps. Again, the basis H2 is a striking exception, indicating that there may be some condition on the basis which is sufficient to enforce this desired behavior.

The performance of the SDPs as coupling is changed is more consistent. The left panel of Figure~\ref{fig:convergence} shows the error in the estimate of the ground-state energy at $\mu = 0.5$, for a sequence of theories of increasing coupling $g^2$. All computations are done on a $10$-site lattice, with the bare mass $m$ tuned such that the renormalized fermion mass (measured on the same lattice) is $m_F = 0.2$. As the coupling is increased, the SDP's error increases roughly linearly.

All bases considered give an exact bound for the free theory. For this theory, the raising and lowering operators that diagonalize the Hamiltonian can be constructed via the discrete Fourier transform of the position-space creation and annihilation operators $c^\dagger(x),c(x)$. Therefore, these operators diagonalizing the Hamiltonian are in the span even of the smallest basis, H0. This is sufficient for the SDP estimate to be exact.

%%%%%%%%%%%%%%%%%%%%%%%%%%%%%%%%%%%%%%%%%%%%%%%%%%%%%%%
\section{Infinite volume}\label{sec:infinite}

As noted in~\cite{barthel2012solving}, because the semidefinite programs constraining expectation values work in terms of operators rather than configurations or states in Hilbert space, these calculations can be performed directly in the infinite volume limit. Of course, even for a system (like the Thirring model) with a finite-dimensional local Hilbert space, the set of operators available in this limit is infinite. As a result, any finite basis chosen to construct the SDP will yield only an approximation. Nevertheless, each estimate of the ground-state energy density will be a rigorous lower bound to the infinite-volume result, and we may hope that the estimates of the number density provide a good approximation in practice.

In the infinite-volume limit, it is no longer possible to speak of the expectation value of the Hamiltonian. Instead, we must work with a Hamiltonian density $\hat h(0)$, defined so that the Hamiltonian is $\hat H = \sum_r \hat h(2r)$:
\begin{widetext}
\begin{eqnarray}
\hat h(x)
&=&
m (c^\dagger(x) c(x) - c^\dagger(x+1) c(x+1))
+ \frac 1 2 (c^\dagger(x) c(x+1) + \mathrm{h.c.})
- \frac 1 2 (c^\dagger(x+1) c(x+2) + \mathrm{h.c.})\nonumber\\
&-& g^2 \big(
c(x) c^\dagger(x) c^\dagger(x+1) c(x+1)
c(x+1) c^\dagger(x+1) c^\dagger(x+2) c(x+2)
\big)
- \mu (c^\dagger(x) c(x) + c^\dagger(x+1)c(x+1))
\text.
\end{eqnarray}
\end{widetext}

The optimization objective for the SDP will be the expectation value $\langle \hat h(0)\rangle$. In order to close out the SDP, we must include some information about how the energy density around site $x=0$ is related to the energy density elsewhere on the lattice. In this case we will assume translational invariance. In particular, for any operator $\mathcal O$ and even displacement $2r$ ($r \in \mathbb Z$), we assume that
\begin{equation}
\langle \mathcal O(x) \rangle = \langle \mathcal O(x+2r) \rangle
\text.
\end{equation}
All linear constraints of this form are added to the SDP.

\begin{figure}
\centering
\includegraphics[width=0.9\linewidth]{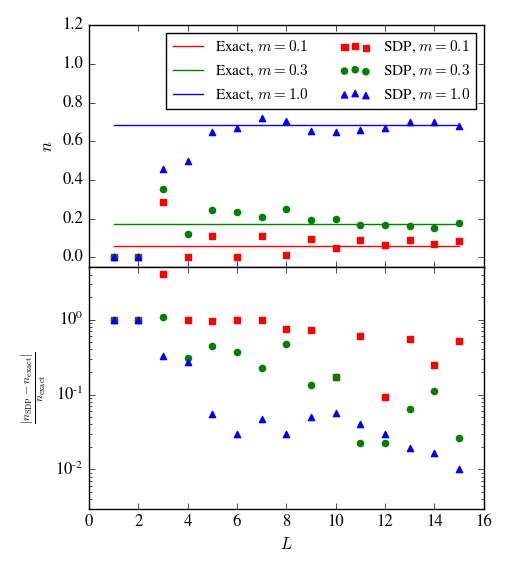}
\caption{Number density estimated via SDPs for free theories. The basis used to construct the SDP, described in the text, grows linearly with the parameter $L$. For all bare masses the SDP converges to the exact number density, but the convergence is somewhat faster for larger masses, where only short-distance correlations are relevant. All calculations are performed at $\frac\mu m = \frac 4 3$. To make the differences in convergence more apparent, the bottom panel shows the relative error on a log scale.\label{fig:free}}
\end{figure}

At finite volume, the free theory is exactly solved by an SDP constructed from a basis consisting only of all available creation and annihilation operators. At infinite volume that basis already must be truncated in order to obtain a finite SDP. Any such truncation will have some maximum radius of operators $L$, and will be an approximation that improves as $L$ is taken to be larger.

The difference between the truncated SDP and the exact result can be thought of as a finite-volume effect. Consider an SDP constructed from the operators $a(-L),\ldots,a(L)$ and their complex conjugates---or, in the case of a theory at non-zero coupling, an SDP consisting of all possible operators with support only on $[-L,L]$. Such an SDP captures all quantum mechanical constraints on expectation values that arise on that range, but has no information about the boundary conditions. As a result, the obtained bound on the energy density can only be as tight as the lowest energy density achievable with any boundary conditions.

The performance of ever-larger SDPs for the free theory is shown in Figure~\ref{fig:free}. Each SDP is constructed only from the operators $a(-L),\ldots,a(L+1)$, and their complex conjugates. The SDP-estimated number density is compared to the exact lattice result
\begin{equation}
n_{\mathrm{exact}}(\mu) = \frac 2 \pi \sin^{-1}\sqrt{\mu^2 - m^2}
\text.
\end{equation}
The most notable feature is that the approximation converges much faster, as a function of $L$, when the fermion mass is larger. Physically, this is a result of the fact that finite volume effects decay parameterically with the dimensionless $mL$.

\begin{figure*}
\centering{
\includegraphics[width=0.45\linewidth]{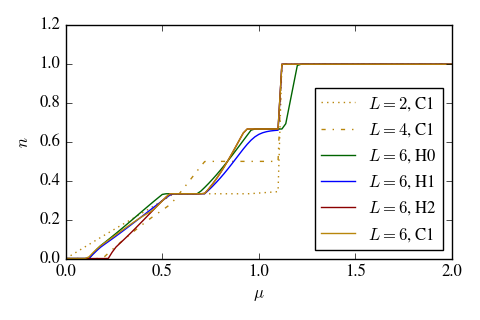}
\hfil
\includegraphics[width=0.45\linewidth]{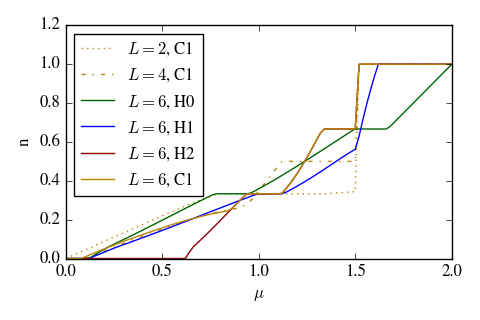}
}
\caption{Number density as a function of chemical potential in the Thirring model, in the infinite volume limit, at weak ($g^2=0.1$, left) and strong ($g^2=0.5$, right) couplings. Both are simulated at a bare fermion mass $m=0.05$. Note that the qualitative agreement between different approximations improves as the basis of operators is enlarged.\label{fig:infinite}}
\end{figure*}

To perform calculations with a nonzero interaction, it is necessary to return to larger bases of the sort considered in the previous two sections. Infinite-volume equivalents of the bases H0, H1, H2, and C1 are defined by simply requiring $-L \le x \le L+1$. Figure~\ref{fig:infinite} shows calculations performed in the infinite volume limit for interacting theories. The left panel shows a weakly coupled theory ($g^2 = 0.1$), and the right a strongly coupled theory ($g^2 = 0.5$).

In both theories, the number of visible plateaus increases as the permitted separation $L$ of operators is increased. This again shows the manner in which the finite-$L$ effects are analogous to finite-volume effects: calculations at finite volume will also show a number of plateaus proportional to the volume of the system.

The weakly-coupled theory in particular shows good qualitative agreement between the various approximations. Although no extrapolation is performed here, both show signs of convergence as larger bases are considered.

%%%%%%%%%%%%%%%%%%%%%%%%%%%%%%%%%%%%%%%%%%%%%%%%%%%%%%%
\section{Discussion}\label{sec:discussion}
This paper has explored the performance of SDP-based methods in analyzing the Thirring model in two spacetime dimensions, at finite fermion density. The SDPs used are small relative to the size of the Hilbert space of the system under study, and there is no indication that the method encounters any obstacles analogous to a fermion sign problem. While Monte Carlo methods perform best at $\mu=0$ and worst when $\mu$ is large enough for the lattice to be saturated, the SDPs have the reverse behavior, with tight bounds most easily achieved for chemical potentials at or near lattice saturation density. 

Fermion sign problems are most often alleviated while remaining within a Monte Carlo framework. It is difficult to construct a fair performance comparison between SDPs and these Monte Carlo-based methods. For example, if we consider the asymptotics of the algorithm when the answer is required to be accurate to within $\epsilon$, we see that SDP methods formally outperform Monte Carlo methods. For a system with a finite-dimensional Hilbert space, a Monte Carlo method requires time exponential in the requested number of significant figures. Since, for a fixed finite-dimensional system, there is an SDP that obtains the exact answer, the time required for the SDP to perform a computation to within that precision is of order $1$, formally\footnote{Strictly speaking, there is a term logarithmic in $\epsilon$ to account for floating-point precision.}---but that observation provides us with no practical information about the performance of SDPs.

The SDPs considered here provide qualitatively accurate equations of state at finite volumes, across the entire range of chemical potentials. However, the equations of state obtained fail a few important tests. Most notably, the estimated number density, and the ``silver blaze'' phenomenon at small chemical potentials is not correctly reproduced.

As with other methods along the lines of the ``quantum-mechanical bootstrap''~\cite{Han:2020bkb,Berenstein:2021dyf,Berenstein:2021loy,Nancarrow:2022wdr,Lawrence:2021msm}, the method in this paper is in a sense dual to the usual variational principle. Whereas a variational study of a quantum-mechanical system yields a rigorous upper bound on the energy, the solution to an SDP yields a rigorous lower bound to the same quantity. As a result, and as demonstrated explicitly in~\cite{Lawrence:2021msm}, the two may be profitably used in conjunction to yield both rigorous bounds on the energy, and a heuristic estimate of how well either algorithm is doing at approximating the true ground state.

In the course of this work, the main obstacle encountered to using larger SDPs was the memory required to store and manipulate the SDP, rather than the time required during solving. The largest SDP considered was based on an $81\times 81$ positive semi-definite matrix, requiring around 2 GB and 90 seconds\footnote{Timed on a 4-core AMD Ryzen 3 5300U.} to solve. It has been noted before~\cite{Lawrence:2021msm} that improved software might enable larger bases to be used; indeed, major advances in the study of the conformal bootstrap over the last decade were made possible by the development of a specialized software package optimized for the sorts of convex optimization problems encountered in that context~\cite{Simmons-Duffin:2015qma}. The fact that the results in this paper are constrained more by memory than by time---and that the memory use is dominated by bookkeeping for the interior point method, rather than the size of the matrix itself---suggests that similar advances in the quantum mechanical bootstrap might be achieved by making a different memory-time tradeoff.

It is also possible to work with dramatically larger SDPs by obtaining only an approximate solution~\cite{burer2003nonlinear,yurtsever2021scalable}. This sacrifices the rigor of the obtained lower bound. However, for field theory calculations the lower bound on the energy is not physically relevant, potentially making this a worthwhile tradeoff.

SDP-based methods in the infinite volume limit display a phenomenon analogous to finite-volume effects due to the finite spatial extent of operators used in the basis. This suggests that standard techniques for extrapolating to the infinite-volume limit may be helpful in improving the accuracy of the approximation. This is particularly applicable to field theories, where the fact that an SDP yields a \emph{bound} on the energy (rather than just an estimate) is not particularly helpful.

This work has indicated that SDPs may represent a practical approach to analyzing the ground-state behavior of fermions at finite density. The application of SDPs to pure-gauge systems has been explored in~\cite{Anderson:2016rcw} (although in the path-integral, rather than Hamiltonian, formalism). A natural next step, therefore, is to test these methods on a gauge theory with a finite fermionic chemical potential.

\begin{acknowledgments}
I am grateful to Paul Romatschke for a variety of helpful discussions over the course of this work, and to Paulo Bedaque, Frederic Koehler, Henry Lamm, and Julian Lenz for comments on drafts of this manuscript. This material is based upon work supported by the U.S. Department of Energy, Office of Science, Office of Nuclear Physics program under Award Number DE-SC-0017905.
\end{acknowledgments}
\bibliographystyle{apsrev4-2}
\bibliography{References}
\end{document}